\documentstyle [12pt,bezier,epsfig]{article}
\newcommand{\GEV}{\mbox{$\mathrm{GeV}$}}
\newcommand{\GEVc}{\mbox{$\mathrm{GeV}/{\it c}$}}

\newcommand{\GeV}{\mbox{GeV}}
\newcommand{\GeVcsq}{\mbox{$\GeV/c^2$}}

\def\ra{\rightarrow} 
\def\gappeq{\mathrel{ \rlap{\raise.5ex\hbox{$>$}}
                      {\lower.5ex\hbox{$\sim$}}  } }
\def\lappeq{\mathrel{ \rlap{\raise.5ex\hbox{$<$}}
                      {\lower.5ex\hbox{$\sim$}}  } }
   
\def\be{\begin{equation}}
\def\ee{\end{equation}}
\def\bea{\begin{eqnarray}}
\def\eea{\end{eqnarray}}
   \def\ra{\rightarrow}

  \def\eqref#1{(\ref{#1})}

   \def\su2{{{\mathrm{ SU(2)}}}_{{\mathrm L}} \times{{\mathrm U(1)}}}

  \def\mf2{m_f^2}

  \def\s0h{{\sigma}^{{\mathrm{ peak,0}}}_{{\mathrm{ had}} }}

  \def\GeV{{\mathrm{GeV}}}

\begin{document}
\begin{titlepage}
\begin{center}
\begin{large}
EUROPEAN LABORATORY FOR PARTICLE PHYSICS (CERN)
\end{large}
\end{center}
\vglue 1.0cm
    \begin{flushright}
        CERN-EP/99-092 \\
        8 July 1999 \\
        \vspace{0.5cm}
    \end{flushright}

\vspace{0.5cm}

\begin{center}
{\huge {\bf  Determination of the\\ 
          LEP centre-of-mass energy \\
           from $Z\gamma$ events } }
\end{center}

\vspace{1cm}

\begin{center}
 {\large  {\bf The ALEPH Collaboration }} 
\end{center}
\vspace{2cm}

\begin{center}
{\bf   Abstract}
\end{center}
 Radiative returns to the $Z$ resonance ($Z\gamma$ events) are used to
determine the LEP2 centre-of-mass energy 
from the data collected with the ALEPH detector in 1997. 
The average centre-of-mass energy is measured to be:
$$E_{CM} = 182.50~\pm~0.19~(stat.)~\pm~0.08 ~(syst.)~\GeV$$
in good agreement with the precise determination by the LEP energy
working group of $182.652\pm0.050~\GeV$. If applied to the measurement of the $W$ mass, 
its precision translates into a  systematic error on $M_W$
which is smaller than the statistical error  achieved 
from the corresponding dataset.

\vspace{2cm}
\begin{center}
{ \em Submitted to Physics Letters B}\\
%
%
\end{center}
\end{titlepage}
%
\pagestyle{empty}
\newpage
\small
%
%
\newlength{\saveparskip}
\newlength{\savetextheight}
\newlength{\savetopmargin}
\newlength{\savetextwidth}
\newlength{\saveoddsidemargin}
\newlength{\savetopsep}
\setlength{\saveparskip}{\parskip}
\setlength{\savetextheight}{\textheight}
\setlength{\savetopmargin}{\topmargin}
\setlength{\savetextwidth}{\textwidth}
\setlength{\saveoddsidemargin}{\oddsidemargin}
\setlength{\savetopsep}{\topsep}
%
%
\setlength{\parskip}{0.0cm}
\setlength{\textheight}{25.0cm}
\setlength{\topmargin}{-1.5cm}
\setlength{\textwidth}{16 cm}
\setlength{\oddsidemargin}{-0.0cm}
\setlength{\topsep}{1mm}
\pretolerance=10000
\centerline{\large\bf The ALEPH Collaboration}
\footnotesize
\vspace{0.5cm}
{\raggedbottom
\begin{sloppypar}
\samepage\noindent
R.~Barate,
D.~Decamp,
P.~Ghez,
C.~Goy,
S.~Jezequel,
J.-P.~Lees,
F.~Martin,
E.~Merle,
\mbox{M.-N.~Minard},
B.~Pietrzyk,
H.~Przysiezniak
\nopagebreak
\begin{center}
\parbox{15.5cm}{\sl\samepage
Laboratoire de Physique des Particules (LAPP), IN$^{2}$P$^{3}$-CNRS,
F-74019 Annecy-le-Vieux Cedex, France}
\end{center}\end{sloppypar}
\vspace{2mm}
\begin{sloppypar}
\noindent
R.~Alemany,
M.P.~Casado,
M.~Chmeissani,
J.M.~Crespo,
E.~Fernandez,
M.~Fernandez-Bosman,
Ll.~Garrido,$^{15}$
E.~Graug\`{e}s,
A.~Juste,
M.~Martinez,
G.~Merino,
R.~Miquel,
Ll.M.~Mir,
P.~Morawitz,
A.~Pacheco,
I.C.~Park,
I.~Riu
\nopagebreak
\begin{center}
\parbox{15.5cm}{\sl\samepage
Institut de F\'{i}sica d'Altes Energies, Universitat Aut\`{o}noma
de Barcelona, 08193 Bellaterra (Barcelona), E-Spain$^{7}$}
\end{center}\end{sloppypar}
\vspace{2mm}
\begin{sloppypar}
\noindent
A.~Colaleo,
D.~Creanza,
M.~de~Palma,
G.~Iaselli,
G.~Maggi,
M.~Maggi,
S.~Nuzzo,
A.~Ranieri,
G.~Raso,
F.~Ruggieri,
G.~Selvaggi,
L.~Silvestris,
P.~Tempesta,
A.~Tricomi,$^{3}$
G.~Zito
\nopagebreak
\begin{center}
\parbox{15.5cm}{\sl\samepage
Dipartimento di Fisica, INFN Sezione di Bari, I-70126 Bari, Italy}
\end{center}\end{sloppypar}
\vspace{2mm}
\begin{sloppypar}
\noindent
X.~Huang,
J.~Lin,
Q. Ouyang,
T.~Wang,
Y.~Xie,
R.~Xu,
S.~Xue,
J.~Zhang,
L.~Zhang,
W.~Zhao
\nopagebreak
\begin{center}
\parbox{15.5cm}{\sl\samepage
Institute of High-Energy Physics, Academia Sinica, Beijing, The People's
Republic of China$^{8}$}
\end{center}\end{sloppypar}
\vspace{2mm}
\begin{sloppypar}
\noindent
D.~Abbaneo,
U.~Becker,$^{19}$
G.~Boix,$^{6}$
M.~Cattaneo,
F.~Cerutti,
V.~Ciulli,
G.~Dissertori,
H.~Drevermann,
R.W.~Forty,
M.~Frank,
F.~Gianotti,
T.C.~Greening,
A.W.~Halley,
J.B.~Hansen,
J.~Harvey,
P.~Janot,
B.~Jost,
I.~Lehraus,
O.~Leroy,
C.~Loomis,
P.~Maley,
P.~Mato,
A.~Minten,
A.~Moutoussi,
F.~Ranjard,
L.~Rolandi,
D.~Schlatter,
M.~Schmitt,$^{20}$
O.~Schneider,$^{2}$
P.~Spagnolo,
W.~Tejessy,
F.~Teubert,
I.R.~Tomalin,
E.~Tournefier,
A.E.~Wright
\nopagebreak
\begin{center}
\parbox{15.5cm}{\sl\samepage
European Laboratory for Particle Physics (CERN), CH-1211 Geneva 23,
Switzerland}
\end{center}\end{sloppypar}
\vspace{2mm}
\begin{sloppypar}
\noindent
Z.~Ajaltouni,
F.~Badaud
G.~Chazelle,
O.~Deschamps,
S.~Dessagne,
A.~Falvard,
C.~Ferdi,
\linebreak
P.~Gay,
C.~Guicheney,
P.~Henrard,
J.~Jousset,
B.~Michel,
S.~Monteil,
\mbox{J-C.~Montret},
D.~Pallin,
P.~Perret,
F.~Podlyski
\nopagebreak
\begin{center}
\parbox{15.5cm}{\sl\samepage
Laboratoire de Physique Corpusculaire, Universit\'e Blaise Pascal,
IN$^{2}$P$^{3}$-CNRS, Clermont-Ferrand, F-63177 Aubi\`{e}re, France}
\end{center}\end{sloppypar}
\vspace{2mm}
\begin{sloppypar}
\noindent
J.D.~Hansen,
J.R.~Hansen,
P.H.~Hansen,
B.S.~Nilsson,
B.~Rensch,
A.~W\"a\"an\"anen
\nopagebreak
\begin{center}
\parbox{15.5cm}{\sl\samepage
Niels Bohr Institute, 2100 Copenhagen, DK-Denmark$^{9}$}
\end{center}\end{sloppypar}
\vspace{2mm}
\begin{sloppypar}
\noindent
G.~Daskalakis,
A.~Kyriakis,
C.~Markou,
E.~Simopoulou,
A.~Vayaki
\nopagebreak
\begin{center}
\parbox{15.5cm}{\sl\samepage
Nuclear Research Center Demokritos (NRCD), GR-15310 Attiki, Greece}
\end{center}\end{sloppypar}
\vspace{2mm}
\begin{sloppypar}
\noindent
A.~Blondel,
\mbox{J.-C.~Brient},
F.~Machefert,
A.~Roug\'{e},
M.~Swynghedauw,
R.~Tanaka,
A.~Valassi,$^{23}$
\linebreak
H.~Videau
\nopagebreak
\begin{center}
\parbox{15.5cm}{\sl\samepage
Laboratoire de Physique Nucl\'eaire et des Hautes Energies, Ecole
Polytechnique, IN$^{2}$P$^{3}$-CNRS, \mbox{F-91128} Palaiseau Cedex, France}
\end{center}\end{sloppypar}
\vspace{2mm}
\begin{sloppypar}
\noindent
E.~Focardi,
G.~Parrini,
K.~Zachariadou
\nopagebreak
\begin{center}
\parbox{15.5cm}{\sl\samepage
Dipartimento di Fisica, Universit\`a di Firenze, INFN Sezione di Firenze,
I-50125 Firenze, Italy}
\end{center}\end{sloppypar}
\vspace{2mm}
\begin{sloppypar}
\noindent
R.~Cavanaugh,
M.~Corden,
C.~Georgiopoulos
\nopagebreak
\begin{center}
\parbox{15.5cm}{\sl\samepage
Supercomputer Computations Research Institute,
Florida State University,
Tallahassee, FL 32306-4052, USA $^{13,14}$}
\end{center}\end{sloppypar}
\vspace{2mm}
\begin{sloppypar}
\noindent
A.~Antonelli,
G.~Bencivenni,
G.~Bologna,$^{4}$
F.~Bossi,
P.~Campana,
G.~Capon,
V.~Chiarella,
P.~Laurelli,
G.~Mannocchi,$^{1,5}$
F.~Murtas,
G.P.~Murtas,
L.~Passalacqua,
M.~Pepe-Altarelli$^{1}$
\nopagebreak
\begin{center}
\parbox{15.5cm}{\sl\samepage
Laboratori Nazionali dell'INFN (LNF-INFN), I-00044 Frascati, Italy}
\end{center}\end{sloppypar}
\vspace{2mm}
\begin{sloppypar}
\noindent
M.~Chalmers,
L.~Curtis,
J.G.~Lynch,
P.~Negus,
V.~O'Shea,
B.~Raeven,
C.~Raine,
D.~Smith,
P.~Teixeira-Dias,
A.S.~Thompson,
J.J.~Ward
\nopagebreak
\begin{center}
\parbox{15.5cm}{\sl\samepage
Department of Physics and Astronomy, University of Glasgow, Glasgow G12
8QQ,United Kingdom$^{10}$}
\end{center}\end{sloppypar}
\begin{sloppypar}
\noindent
O.~Buchm\"uller,
S.~Dhamotharan,
C.~Geweniger,
P.~Hanke,
G.~Hansper,
V.~Hepp,
E.E.~Kluge,
A.~Putzer,
J.~Sommer,
K.~Tittel,
S.~Werner,$^{19}$
M.~Wunsch
\nopagebreak
\begin{center}
\parbox{15.5cm}{\sl\samepage
Institut f\"ur Hochenergiephysik, Universit\"at Heidelberg, D-69120
Heidelberg, Germany$^{16}$}
\end{center}\end{sloppypar}
\vspace{2mm}
\begin{sloppypar}
\noindent
R.~Beuselinck,
D.M.~Binnie,
W.~Cameron,
P.J.~Dornan,$^{1}$
M.~Girone,
S.~Goodsir,
N.~Marinelli,
E.B.~Martin,
J.~Nash,
J.~Nowell,
A.~Sciab\`a,
J.K.~Sedgbeer,
E.~Thomson,
M.D.~Williams
\nopagebreak
\begin{center}
\parbox{15.5cm}{\sl\samepage
Department of Physics, Imperial College, London SW7 2BZ,
United Kingdom$^{10}$}
\end{center}\end{sloppypar}
\vspace{2mm}
\begin{sloppypar}
\noindent
V.M.~Ghete,
P.~Girtler,
E.~Kneringer,
D.~Kuhn,
G.~Rudolph
\nopagebreak
\begin{center}
\parbox{15.5cm}{\sl\samepage
Institut f\"ur Experimentalphysik, Universit\"at Innsbruck, A-6020
Innsbruck, Austria$^{18}$}
\end{center}\end{sloppypar}
\vspace{2mm}
\begin{sloppypar}
\noindent
C.K.~Bowdery,
P.G.~Buck,
G.~Ellis,
A.J.~Finch,
F.~Foster,
G.~Hughes,
R.W.L.~Jones,
N.A.~Robertson,
M.~Smizanska,
M.I.~Williams
\nopagebreak
\begin{center}
\parbox{15.5cm}{\sl\samepage
Department of Physics, University of Lancaster, Lancaster LA1 4YB,
United Kingdom$^{10}$}
\end{center}\end{sloppypar}
\vspace{2mm}
\begin{sloppypar}
\noindent
I.~Giehl,
F.~H\"olldorfer,
K.~Jakobs,
K.~Kleinknecht,
M.~Kr\"ocker,
A.-S.~M\"uller,
H.-A.~N\"urnberger,
G.~Quast,
B.~Renk,
E.~Rohne,
H.-G.~Sander,
S.~Schmeling,
H.~Wachsmuth
C.~Zeitnitz,
T.~Ziegler
\nopagebreak
\begin{center}
\parbox{15.5cm}{\sl\samepage
Institut f\"ur Physik, Universit\"at Mainz, D-55099 Mainz, Germany$^{16}$}
\end{center}\end{sloppypar}
\vspace{2mm}
\begin{sloppypar}
\noindent
J.J.~Aubert,
C.~Benchouk,
A.~Bonissent,
J.~Carr,$^{1}$
P.~Coyle,
A.~Ealet,
D.~Fouchez,
F.~Motsch,
P.~Payre,
D.~Rousseau,
M.~Talby,
M.~Thulasidas,
A.~Tilquin
\nopagebreak
\begin{center}
\parbox{15.5cm}{\sl\samepage
Centre de Physique des Particules, Facult\'e des Sciences de Luminy,
IN$^{2}$P$^{3}$-CNRS, F-13288 Marseille, France}
\end{center}\end{sloppypar}
\vspace{2mm}
\begin{sloppypar}
\noindent
M.~Aleppo,
M.~Antonelli,
S.~Gilardoni,
F.~Ragusa
\nopagebreak
\begin{center}
\parbox{15.5cm}{\sl\samepage
Dipartimento di Fisica, Universit\`a di Milano e INFN Sezione di
Milano, I-20133 Milano, Italy.}
\end{center}\end{sloppypar}
\vspace{2mm}
\begin{sloppypar}
\noindent
V.~B\"uscher,
H.~Dietl,
G.~Ganis,
K.~H\"uttmann,
G.~L\"utjens,
C.~Mannert,
W.~M\"anner,
\mbox{H.-G.~Moser},
S.~Schael,
R.~Settles,
H.~Seywerd,
H.~Stenzel,
W.~Wiedenmann,
G.~Wolf
\nopagebreak
\begin{center}
\parbox{15.5cm}{\sl\samepage
Max-Planck-Institut f\"ur Physik, Werner-Heisenberg-Institut,
D-80805 M\"unchen, Germany\footnotemark[16]}
\end{center}\end{sloppypar}
\vspace{2mm}
\begin{sloppypar}
\noindent
P.~Azzurri,
J.~Boucrot,
O.~Callot,
S.~Chen,
M.~Davier,
L.~Duflot,
\mbox{J.-F.~Grivaz},
Ph.~Heusse,
A.~Jacholkowska,$^{1}$
M.~Kado,
J.~Lefran\c{c}ois,
L.~Serin,
\mbox{J.-J.~Veillet},
I.~Videau,$^{1}$
J.-B.~de~Vivie~de~R\'egie,
D.~Zerwas
\nopagebreak
\begin{center}
\parbox{15.5cm}{\sl\samepage
Laboratoire de l'Acc\'el\'erateur Lin\'eaire, Universit\'e de Paris-Sud,
IN$^{2}$P$^{3}$-CNRS, F-91898 Orsay Cedex, France}
\end{center}\end{sloppypar}
\vspace{2mm}
\begin{sloppypar}
\noindent
G.~Bagliesi,
S.~Bettarini,
T.~Boccali,
C.~Bozzi,$^{12}$
G.~Calderini,
R.~Dell'Orso,
I.~Ferrante,
A.~Giassi,
A.~Gregorio,
F.~Ligabue,
A.~Lusiani,
P.S.~Marrocchesi,
A.~Messineo,
F.~Palla,
G.~Rizzo,
G.~Sanguinetti,
G.~Sguazzoni,
R.~Tenchini,
C.~Vannini,
A.~Venturi,
P.G.~Verdini
\samepage
\begin{center}
\parbox{15.5cm}{\sl\samepage
Dipartimento di Fisica dell'Universit\`a, INFN Sezione di Pisa,
e Scuola Normale Superiore, I-56010 Pisa, Italy}
\end{center}\end{sloppypar}
\vspace{2mm}
\begin{sloppypar}
\noindent
G.A.~Blair,
J.~Coles,
G.~Cowan,
M.G.~Green,
D.E.~Hutchcroft,
L.T.~Jones,
T.~Medcalf,
J.A.~Strong,
J.H.~von~Wimmersperg-Toeller
\nopagebreak
\begin{center}
\parbox{15.5cm}{\sl\samepage
Department of Physics, Royal Holloway \& Bedford New College,
University of London, Surrey TW20 OEX, United Kingdom$^{10}$}
\end{center}\end{sloppypar}
\vspace{2mm}
\begin{sloppypar}
\noindent
D.R.~Botterill,
R.W.~Clifft,
T.R.~Edgecock,
P.R.~Norton,
J.C.~Thompson
\nopagebreak
\begin{center}
\parbox{15.5cm}{\sl\samepage
Particle Physics Dept., Rutherford Appleton Laboratory,
Chilton, Didcot, Oxon OX11 OQX, United Kingdom$^{10}$}
\end{center}\end{sloppypar}
\vspace{2mm}
\begin{sloppypar}
\noindent
\mbox{B.~Bloch-Devaux},
P.~Colas,
B.~Fabbro,
G.~Fa\"if,
E.~Lan\c{c}on,
\mbox{M.-C.~Lemaire},
E.~Locci,
P.~Perez,
J.~Rander,
\mbox{J.-F.~Renardy},
A.~Rosowsky,
A.~Trabelsi,$^{21}$
B.~Tuchming,
B.~Vallage
\nopagebreak
\begin{center}
\parbox{15.5cm}{\sl\samepage
CEA, DAPNIA/Service de Physique des Particules,
CE-Saclay, F-91191 Gif-sur-Yvette Cedex, France$^{17}$}
\end{center}\end{sloppypar}
\vspace{2mm}
\begin{sloppypar}
\noindent
S.N.~Black,
J.H.~Dann,
H.Y.~Kim,
N.~Konstantinidis,
A.M.~Litke,
M.A. McNeil,
G.~Taylor
\nopagebreak
\begin{center}
\parbox{15.5cm}{\sl\samepage
Institute for Particle Physics, University of California at
Santa Cruz, Santa Cruz, CA 95064, USA$^{22}$}
\end{center}\end{sloppypar}
\vspace{2mm}
\begin{sloppypar}
\noindent
C.N.~Booth,
S.~Cartwright,
F.~Combley,
P.N.~Hodgson,
M.S.~Kelly,
M.~Lehto,
L.F.~Thompson
\nopagebreak
\begin{center}
\parbox{15.5cm}{\sl\samepage
Department of Physics, University of Sheffield, Sheffield S3 7RH,
United Kingdom$^{10}$}
\end{center}\end{sloppypar}
\vspace{2mm}
\begin{sloppypar}
\noindent
K.~Affholderbach,
A.~B\"ohrer,
S.~Brandt,
C.~Grupen,
J.~Hess,
A.~Misiejuk,
G.~Prange,
U.~Sieler
\nopagebreak
\begin{center}
\parbox{15.5cm}{\sl\samepage
Fachbereich Physik, Universit\"at Siegen, D-57068 Siegen, Germany$^{16}$}
\end{center}\end{sloppypar}
\vspace{2mm}
\begin{sloppypar}
\noindent
G.~Giannini,
B.~Gobbo
\nopagebreak
\begin{center}
\parbox{15.5cm}{\sl\samepage
Dipartimento di Fisica, Universit\`a di Trieste e INFN Sezione di Trieste,
I-34127 Trieste, Italy}
\end{center}\end{sloppypar}
\vspace{2mm}
\begin{sloppypar}
\noindent
J.~Putz,
J.~Rothberg,
S.~Wasserbaech,
R.W.~Williams
\nopagebreak
\begin{center}
\parbox{15.5cm}{\sl\samepage
Experimental Elementary Particle Physics, University of Washington, WA 98195
Seattle, U.S.A.}
\end{center}\end{sloppypar}
\vspace{2mm}
\begin{sloppypar}
\noindent
S.R.~Armstrong,
P.~Elmer,
D.P.S.~Ferguson,
Y.~Gao,
S.~Gonz\'{a}lez,
O.J.~Hayes,
H.~Hu,
S.~Jin,
P.A.~McNamara III,
J.~Nielsen,
W.~Orejudos,
Y.B.~Pan,
Y.~Saadi,
I.J.~Scott,
J.~Walsh,
Sau~Lan~Wu,
X.~Wu,
G.~Zobernig
\nopagebreak
\begin{center}
\parbox{15.5cm}{\sl\samepage
Department of Physics, University of Wisconsin, Madison, WI 53706,
USA$^{11}$}
\end{center}\end{sloppypar}
}
\footnotetext[1]{Also at CERN, 1211 Geneva 23, Switzerland.}
\footnotetext[2]{Now at Universit\'e de Lausanne, 1015 Lausanne, Switzerland.}
\footnotetext[3]{Also at Centro Siciliano di Fisica Nucleare e Struttura
della Materia, INFN Sezione di Catania, 95129 Catania, Italy.}
\footnotetext[4]{Also Istituto di Fisica Generale, Universit\`{a} di
Torino, 10125 Torino, Italy.}
\footnotetext[5]{Also Istituto di Cosmo-Geofisica del C.N.R., Torino,
Italy.}
\footnotetext[6]{Supported by the Commission of the European Communities,
contract ERBFMBICT982894.}
\footnotetext[7]{Supported by CICYT, Spain.}
\footnotetext[8]{Supported by the National Science Foundation of China.}
\footnotetext[9]{Supported by the Danish Natural Science Research Council.}
\footnotetext[10]{Supported by the UK Particle Physics and Astronomy Research
Council.}
\footnotetext[11]{Supported by the US Department of Energy, grant
DE-FG0295-ER40896.}
\footnotetext[12]{Now at INFN Sezione di Ferrara, 44100 Ferrara, Italy.}
\footnotetext[13]{Supported by the US Department of Energy, contract
DE-FG05-92ER40742.}
\footnotetext[14]{Supported by the US Department of Energy, contract
DE-FC05-85ER250000.}
\footnotetext[15]{Permanent address: Universitat de Barcelona, 08208 Barcelona,
Spain.}
\footnotetext[16]{Supported by the Bundesministerium f\"ur Bildung,
Wissenschaft, Forschung und Technologie, Germany.}
\footnotetext[17]{Supported by the Direction des Sciences de la
Mati\`ere, C.E.A.}
\footnotetext[18]{Supported by Fonds zur F\"orderung der wissenschaftlichen
Forschung, Austria.}
\footnotetext[19]{Now at SAP AG, 69185 Walldorf, Germany}
\footnotetext[20]{Now at Harvard University, Cambridge, MA 02138, U.S.A.}
\footnotetext[21]{Now at D\'epartement de Physique, Facult\'e des Sciences de Tunis, 1060 Le Belv\'ed\`ere, Tunisia.}
\footnotetext[22]{Supported by the US Department of Energy,
grant DE-FG03-92ER40689.}
\footnotetext[23]{Now at LAL, 91898 Orsay, France.}
%
%
\setlength{\parskip}{\saveparskip}
\setlength{\textheight}{\savetextheight}
\setlength{\topmargin}{\savetopmargin}
\setlength{\textwidth}{\savetextwidth}
\setlength{\oddsidemargin}{\saveoddsidemargin}
\setlength{\topsep}{\savetopsep}
\normalsize
\newpage
\pagestyle{plain}
\setcounter{page}{1}

\clearpage
\section{Introduction}
\label{intro}

One of the main goals of LEP2 is the direct measurement of the $W$ mass
with an accuracy  better than 30 MeV. 
This requires a very 
precise determination of the LEP centre-of-mass energy, in order to
minimize its contribution to the $W$ mass uncertainty.  
At LEP1 this accuracy was achieved 
using the Resonant 
Depolarisation  method for energy measurement~\cite{lep1cal}.
At LEP2 this is no longer possible 
since polarization has not been achieved above a beam energy of 60 GeV. 
Therefore, the precise Resonant Depolarisation calibration is performed only 
in the 40-55~GeV energy range,
and  extrapolation to higher energies is necessary.
The  precision achieved on the 183~GeV 
centre-of-mass energy range is about 
$\pm 50$~MeV~\cite{lepewg}. For future data taking there is hope to improve this precision
using the recently installed magnetic spectrometer~\cite{weblepewg}.

   An alternative determination  is considered in this paper.
 It is based on an analysis of 
$e^+e^-\rightarrow \gamma f\bar{f}$ events, where the Initial State
Radiation (ISR) photons are mostly collinear to the beam pipe and remain 
undetected.  In these events, 
the invariant mass, $\sqrt{s'}$, of the final state $f\bar{f}$ system 
peaks naturally at the $Z$ mass. 
The variable  $x= 1- {s'/s}$ can be reconstructed  
from the final state particle angles 
which are well measured quantities. 
Since the $Z$ mass is very well known from LEP1, 
a fit to the  distribution of $x$ 
can be used to determine the center-of-mass energy
$\sqrt{s}$.

This letter presents an application of this method using the $q\bar{q}$ final state in data recorded by ALEPH in 1997.
LEP ran at  nominal centre-of-mass energies  
of 181, 182, 183 and 184 GeV, and an integrated luminosity of 56.812 pb$^{-1}$
was collected.

\section{The ALEPH detector}
\label{sec:detector}

A detailed description of the ALEPH detector can be found in
Ref.~\cite{det}
and of its performance in Ref.~\cite{EFLW}. Charged particles are
detected in
the central part of the detector. From the beam crossing point outwards, 
a silicon vertex
detector, a cylindrical drift chamber, and a large time projection
chamber (TPC)
measure up to 31 coordinates along the charged particle
trajectories.
A 1.5~T axial magnetic field is provided
by a superconducting solenoidal coil. 
A resolution of 
$\delta p_T/p_T=6 \times 10^{-4} p_T \oplus 0.005$ ($p_T$ in \GEVc)
is achieved. 
Hereafter, charged particle tracks reconstructed 
from at least four hits in the TPC, having 
a polar angle to the beam axis satisfying $|{\cos\theta}|$~$<0.95$,  and
  originating
from within a cylinder of 2~cm radius and 20~cm length, 
centred on the nominal interaction point 
and parallel to the beam axis, are called {\it good tracks}.

Electrons and photons are identified in the electromagnetic calorimeter
by their shower profile. 
The calorimeter, a lead/wire-plane sampling device
with fine readout segmentation and total thickness of 22 radiation 
lengths at normal incidence, provides an energy resolution $\Delta E/E$ 
of $0.180/\sqrt{E}+0.009$ ($E$ in \GEV).

Muons are identified by their penetration pattern in the
hadron 
calorimeter, a 1.2~m thick iron yoke instrumented with 23 layers of streamer
tubes, 
together with two surrounding layers of muon chambers. In association 
with the electromagnetic calorimeter, the hadron
calorimeter also provides a measurement of the  energy of charged and
neutral hadrons with a relative resolution of $0.85/\sqrt{E}$ ($E$ in
\GEV).

The total visible energy and momentum, as well as the missing energy,
are evaluated by an energy flow reconstruction algorithm~\cite{EFLW} 
which combines all of the above measurements, complemented  at low polar
angles by the energy detected in the luminosity 
calorimeters. 
Jets are built from
charged and neutral
 objects reconstructed by the energy flow algorithm.  
The typical jet angular resolution is 30~mrad. 
The jet energy
resolution is approximately
$\sigma_E = (0.6\sqrt{E}+0.6) {\GEV} \cdot (1+\cos^2\theta)$, 
where  $E$ (in \GEV) and $\theta$ are 
the jet energy and polar angle.
The jet energy and angular resolution as well as calibrations were obtained from
extensive studies of 
$Z\ra q\bar{q}$ events both in data and Monte Carlo.
Discrepancies between data and simulation were 
used when evaluating systematic errors.  


\section{Monte Carlo samples}
\label{mc} 
A sample of 200,000 $q\bar{q}$ events 
was generated using  KORALZ~v4.2 \cite{KRLZ} 
at a nominal energy of 182.675 GeV 
and
fully simulated in the detector. 
Monte Carlo samples at 183 GeV with integrated luminosities at least 20 times larger
 than recorded 
were simulated for all background reactions. PYTHIA~v5.7 \cite{PYTH} 
 was used to generate $ZZ$ and $Zee$ events, and KORALW~v1.21 ~\cite{KORALW} 
 to produce four fermion  events with $WW$ topologies. Two-photon ($\gamma\gamma$) 
reactions into hadrons were  simulated with PHOT02 \cite{PHOT} 
and PHOJET \cite{PHOJET}.
%
The method was calibrated with  
fully simulated $q\bar{q}$ samples
of 20,000 events, generated with KORALZ, at 181, 182, 184 and 185 GeV.



\section{Event selection and reconstruction algorithm}
\label{selection}
At $\sqrt{s}$ = 183 GeV the main backgrounds to the 
process $e^+e^- \rightarrow Z \gamma \rightarrow  q \bar{q} \gamma $ are 
$e^+e^- \rightarrow  WW$, $e^+e^- \rightarrow  ZZ$,
$e^+e^- \rightarrow  \gamma\gamma$ and 
$e^+e^- \rightarrow  Zee$. 
To discriminate between signal and background events, 
the following cuts are applied.
\begin{itemize}
\item{}  $q\bar{q}$ candidates are  
                 required to have at least seven good tracks
                 which total 
                 energy exceeds 10\%
                of the nominal centre-of-mass energy.  
\item{} The energy flow objects are clustered into jets using the JADE algorithm~\cite{JADE}
        with a  ${\sl y}_{cut}$ value of 0.008. Events containing jets with an electromagnetic content
        of at least 90\%  and an energy larger than 10 GeV are rejected. 
\item{} The total visible invariant mass is required 
                to be greater than 50~{\GeVcsq}.
\item{} The remaining events are then forced into two
                jets using the DURHAM-E algorithm. The polar angle of both jets, $\theta$, is 
                restricted to $|{\cos\theta}|<0.95$.
\end{itemize} 
To reconstruct the effective 
               centre-of-mass energy $\sqrt{s'}$, 
               it is assumed that the ISR photon is 
               emitted along the beam pipe, resulting in a boost of the $f \bar{f}$ system,
                or the produced $Z$,  
               in the opposite direction. 
               A kinematic reconstruction  
               based on rescaling of the jet energies is performed for each event. 
               According to Monte Carlo studies,
               jet velocities and angles 
              are well measured. Each jet four-momentum is scaled by a factor 
               $\alpha$ according to the relations
               \begin{center}
                \begin{eqnarray}
                  \nonumber
                 E^{rec}_1=\alpha_1 E^{meas}_1 &;& 
                 {\overrightarrow {P_1}}^{rec}=\alpha_1 {\overrightarrow {P_1}}^{meas}\\
                  \nonumber  
                E^{rec}_2=\alpha_2 E^{meas}_2 &;&
                  {\overrightarrow {P_2}}^{rec}=\alpha_2 {\overrightarrow {P_2}}^{meas}
                 \end{eqnarray}
               \end{center}
               where {\it rec} stands for reconstructed and {\it meas} for measured.
               Correction coefficients  are 
               different for the two jets.

               In the absence of final state radiation (FSR), the correction 
               coefficients $\alpha_1$ and $\alpha_2$
               are obtained by applying 
               energy and momentum conservation, and the assumption that
               one zero mass particle escapes detection (the ISR $\gamma$)
               along the beam axis.
               The solution kept is the one 
               which yields   $\alpha_1$ and $\alpha_2$ positive and closest to 1.
               Then, the effective  centre-of-mass energy can be expressed as
 \begin{eqnarray}
   \nonumber
                 s' &=& s~ F~\left(
\frac{{\vec{P}}^{\rm meas}_{1}}{E^{\rm meas}_1} ,
\frac{{\vec{P}}^{\rm meas}_{2}}{E^{\rm meas}_2}
\right)
 \end{eqnarray}
from which the quantity  $x=1-(s'/s)$ is reconstructed independently of the 
nominal centre-of-mass energy.

               The fit to the centre-of-mass energy is performed for events with   
                 0.60 $< x <$ 0.88 where the highest purity and  the 
                 maximum sensitivity is achieved. 
The generated  
cross-sections, the effective ones  after selection cuts 
and within the fitting window  
  are summarised for each process in Table~\ref{efftab}.
\begin{table}[htb]
\begin{center}
\begin{tabular}{|c|c|c|c|}
\hline
Processes &$\sigma$(pb) & $\sigma_{eff}$(pb) & $\sigma_{window}$(pb) \\
\hline
$q\bar{q}\gamma$ & 108.84 $\pm$ 0.08 & 68.61 $\pm$ 0.20  & 40.07 $\pm$ 0.15  \\
\hline
\hline
$WW$ & 16.02 $\pm$ 0.01 & 10.51 $\pm$ 0.05  & 0.47 $\pm$ 0.01  \\
\hline
$ZZ$ & 2.545 $\pm$ 0.13 & 1.08 $\pm$ 0.05  &  0.33 $\pm$ 0.02  \\
\hline
$\gamma\gamma \rightarrow$ $hadrons$ $(untagged)$  & 7800. $\pm$ 390.& 0.98 $\pm$ 0.22 & 0.24 $\pm$ 0.11 \\
\hline
$\gamma\gamma \rightarrow$ $u/d$  & 474.00 $\pm$ 23.70 & 0.61 $\pm$ 0.08  & 0.19 $\pm$ 0.04   \\
\hline
$\gamma\gamma \rightarrow$ $ss$  & 26.00 $\pm$ 1.30 & 0.04 $\pm$ 0.02 & 0.01 $\pm$ 0.01 \\
\hline
$\gamma\gamma \rightarrow$ $cc$  & 93.60 $\pm$ 4.68 & 0.55 $\pm$ 0.008 &  0.19 $\pm$ 0.05\\
\hline
$\gamma\gamma \rightarrow$ $bb$  & 0.53 $\pm$ 0.03 & 0.02 $\pm$ 0.005&  0.01 $\pm$ 0.01 \\
\hline
$Zee$   & 6.80 $\pm$ 0.27 & 4.57 $\pm$ 0.18 & 1.02 $\pm$ 0.04   \\
\hline
$We\nu$ & 0.608 $\pm$ 0.03 & 0.32 $\pm$ 0.02  & 0.11 $\pm$ 0.01  \\
\hline
\hline
\multicolumn{2}{|c|}{Purity (\%)} & 78.60 $\pm$ 0.01 & 93.93 $\pm$ 0.01 \\
\hline
\end{tabular}
\end{center}
\caption[]  
{\protect\footnotesize
Generated (second column) and effective cross-section after selection cuts (third column) and within
 the fitting window (fourth column), for signal and background processes (first column)
 at $\sqrt{s}$=182.675.}
\label{efftab}
\end{table}



\section{LEP Centre-of-Mass Energy measurement}
\label{measurement}

The LEP centre-of-mass energy is determined for each of the 181, 182, 183 and 184~GeV datasets.
The four values are combined  taking into account the relative luminosity of each dataset.
A Monte Carlo  reweighting procedure \cite{WMASS} is
applied to  find the value of E$_{CM}$ which best fits the reconstructed $x$ distribution. 
Selected Monte Carlo (KORALZ) signal events from the large sample at a reference 
energy of 182.675~GeV are reweighted using 
the ratio of the differential production cross-sections \cite{FEDEMISA} 
\begin{eqnarray}
 \nonumber
 {\it w}_i(E_{CM}) & = &
  \frac{\frac{{\it d\sigma}}{{\it dx}}(x_{true}^{i},E_{CM})}
  {\frac{{\it d\sigma}}{{\it dx}}(x_{true}^{i},E^{ref}_{CM})}, 
\end{eqnarray}
where $x_{i}$ denotes the $x$ at which the $i$th event has been generated.  
Background Monte Carlo
 samples are included  in the fit, but they are 
not reweighted, i.e., their energy dependence is not taken into account. This introduces
 a  systematic error which is estimated in section~\ref{backsys}.

 A maximum likelihood fit is performed with fixed
bins of 0.01 over the $x$ range of 0.60--0.88.
The statistical error on the centre-of-mass energy is derived  from the individual fits to the 
data distributions.

The linearity of the reweighting technique was verified using 
five independent Monte Carlo samples (signal and background)
generated  within 2 GeV around the central nominal energy.
The fitted values from these distributions have a
slope 1.035$\pm$0.034, consistent with unity, and an insignificant 
offset of 0.045$\pm$0.049 GeV.

\section{Results}
\label{results}
The number of selected events  
 at each nominal energy, and the corresponding
number of expected events are summarized in Table~\ref{data}. The uncertainty 
on the number of expected events comes from three sources:
the error on the integrated luminosity measurement, the statistical error 
 on the estimation of the efficiencies  and the theoretical
uncertainty on the cross-sections.
\begin{table}[htb]
\begin{center}
\begin{tabular}{|c|c|c|c|}
\hline
Nominal & Data   & Expected &  Integrated \\
dataset & events &  events  &  Luminosity (pb$^{-1}$)\\
\hline
181 &   11  &    8 $\pm$ 1 & 0.166 $\pm$ 0.006 \\
\hline
182 &  187    &  173 $\pm$ 2 & 3.924 $\pm$  0.029 \\
\hline
183 & 2293  & 2191 $\pm$ 12 & 50.795 $\pm$ 0.105 \\
\hline
184  & 76  &   84 $\pm$  1  & 1.927 $\pm$ 0.021  \\
\hline
\end{tabular}
\end{center}
\caption[]  
{\protect\footnotesize Number of events found in  each energy dataset after the 
selection cuts, as well as the corresponding
number of expected events  and  integrated luminosity 
.}
\label{data}
\end{table}

The LEP centre-of-mass energy values and the statistical errors obtained from fits to
the individual datasets are summarized in Table~\ref{energy}, as well as the weighted average
according to the integrated luminosity of each dataset.
The last column gives the value obtained by the LEP energy calibration group.
\begin{table}[htb]
\begin{center}
\begin{tabular}{|c|c|c|c|c|}
\hline
Nominal & E$_{CM}$(fit) & -$\Delta$E$_{CM}$ & +$\Delta$E$_{CM}$ &  E$_{CM}$(LEP)\\
energy & (GeV) & (GeV) & (GeV) & (GeV) \\
\hline
181 & 181.19  & $-$2.45 &  $+$2.82 & 180.826 $\pm$ 0.050 \\
\hline
182 & 181.73  & $-$0.63 & $+$0.68  & 181.708 $\pm$ 0.050 \\
\hline
183 & 182.56  & $-$0.22 & $+$0.19 & 182.691 $\pm$ 0.050 \\
\hline
184  & 182.56 & $-$0.99 & $+$1.11 & 183.801 $\pm$ 0.050\\
\hline
\hline
Combined & 182.50 & $-$0.20 & $+$0.18 & 182.652 $\pm$ 0.050 \\
\hline
\end{tabular}
\end{center}
\caption[]  
{\protect\footnotesize 
Fitted LEP centre-of-mass energies and  corresponding 
statistical errors for each  dataset. 
The last column gives the LEP centre-of-mass energies 
established by the LEP energy calibration group 
for the corresponding dataset.}
\label{energy}
\end{table}

\begin{figure*}[htb!]
\vspace{-1.cm}
\begin{center}
\mbox{\epsfig{file=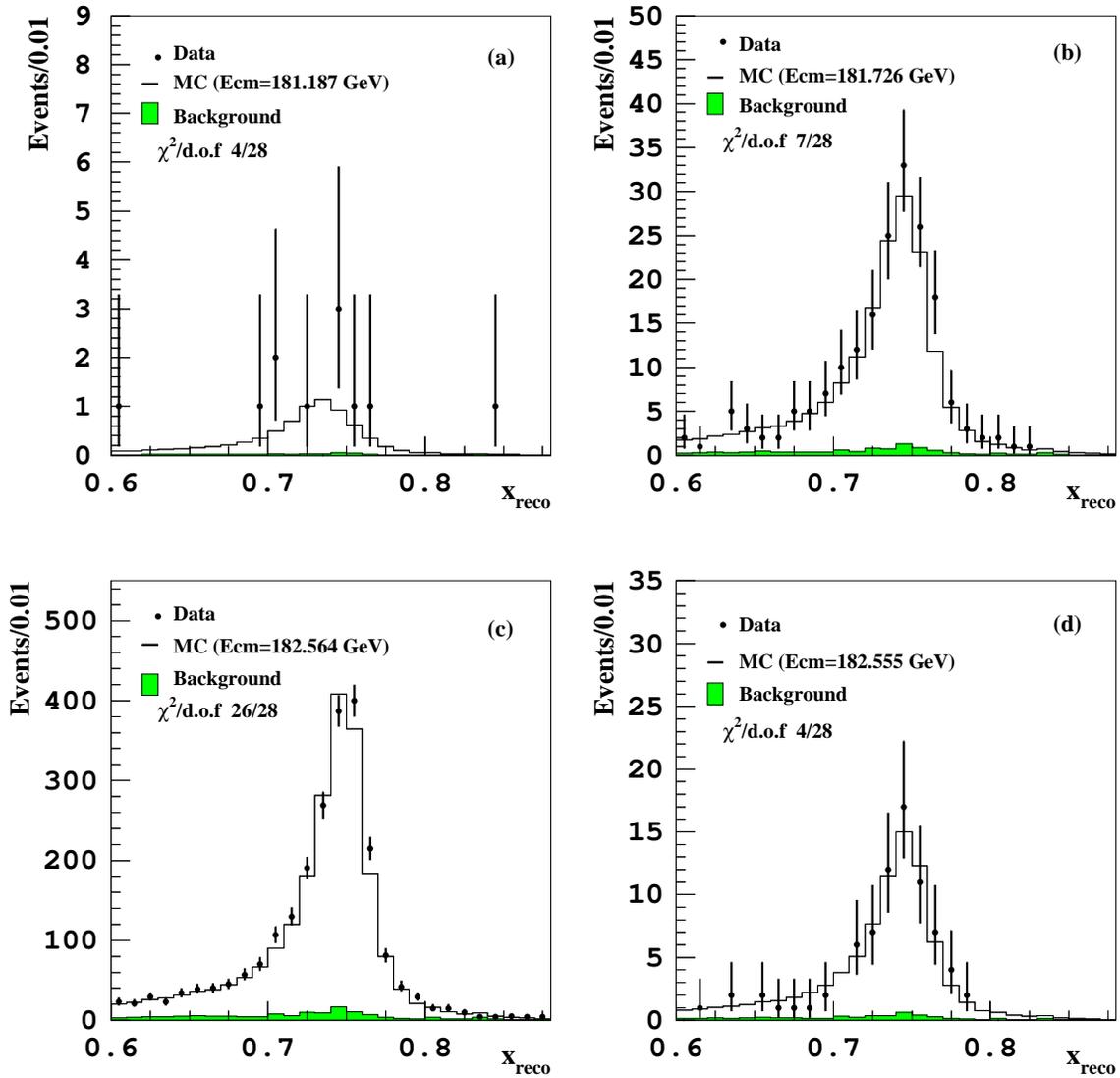,height=17cm}}
\end{center}
\vspace{-1.2cm}
\caption[]
{\protect\footnotesize
 Reconstructed x distribution  for selected data (dots)
and  background (shaded area) for each energy 
dataset: 181 (a), 182 (b), 183 (c) and 184 GeV (d).
The open histogram shows the results of the fit} 
\label{datafitfig}
\end{figure*}

Figure~\ref{datafitfig} shows the reconstructed $x$ distribution for the selected 
events in each dataset, superimposed with the Monte Carlo reweighted expectation
using the LEP centre-of-mass energy which best fits the data. 

\section{Systematic uncertainties}
Table~\ref{syslist} summarizes all the systematic errors discussed below.

\begin{table}[htb]
\begin{center}
\begin{tabular}{|c|c|}
\hline
Source & $\Delta$E$_{CM}$ (MeV) \\
\hline
\hline
Calorimeter  calibrations & 44 \\
\hline
Jet energy calibration   & 47 \\
\hline
Jet angle calibration   & 20  \\
\hline
MC statistics      & 24 \\
\hline
ISR & 21\\
\hline
ISR-FSR interference   & 10 \\
\hline
Background contamination & 32 \\

\hline
\hline
Total &   82\\
\hline
\end{tabular}
\end{center}
\caption[]  
{\protect\footnotesize Summary of the systematic errors on 
the LEP centre-of-mass energy measurement. }
\label{syslist}
\end{table}

\noindent{\bf Calorimeter calibrations}

        During the 1997 data taking the uncertainties on the absolute calibration scale
of ECAL and HCAL energies were estimated to be $\pm$0.9\% and $\pm$2\%, respectively.
Maximum shifts of 26~MeV and 36~MeV were observed when applying the fitting procedure 
to the MC sample
after  variation of ECAL and HCAL energies 
in both directions by the corresponding amount. 

\noindent{\bf Jet energy calibration}

$Z$ peak data are used to map the response of the detector
to hadronic jets as a function of their polar angle. The observed difference
in the energy scale of  jets between data and Monte Carlo is
parametrised as a function of 
 $|{\cos\theta}|$, where $\theta$ is the angle between the jet direction and the beam axis.
To estimate the systematic error, two modified parametrisations are derived
which correspond  to
 $\pm$1$\sigma$ uncertainty on the discrepancy. 
 The largest shift observed when these modified parametrisations are used to correct the jet 
energies amounts to 47~MeV and is taken as the 
systematic error.  

\noindent{\bf Jet angle calibration}

$Z$ peak data have been used as well to study the two jets 
acollinearity as a function of $|{\cos\theta}|$.
The observed difference between data and Monte Carlo is 
used  to estimate this systematic error. A 20 MeV shift is observed 
when applying or not this correction and is quoted as the systematic 
error.

%

\noindent{\bf Initial State Radiation}

ISR is described in KORALZ 
up to second 
order in the leading-log approximation ({\cal O}($\alpha^2L^2$)), 
in the YFS \cite{YFS} style. 
The effect of the missing higher order terms   is 
studied by degrading KORALZ to {\cal O}($\alpha^1L^1$). 
A 21~MeV shift is observed and  is quoted as an upper limit 
of the systematic error.

\noindent{\bf ISR-FSR interference}

No existing Monte Carlo generator is able to give a good representation
 of initial-final state
interference for quark final states. An evaluation of the order of 
magnitude of the effect has been performed  
using  ZFITTER~\cite{ZFIT}. The differential
cross-section {{\it d$\sigma$}}/{{\it dx}}   
has been computed with ZFITTER, switching on and off the ISR-FSR interference,
including only the polar angle region $|{\cos(\theta)}|<$0.95, at the energy corresponding 
to the fitted value in the  data. In
both cases these differential cross-sections are integrated in $x$ bins
of 0.01 over the $x$ range of 0.60--0.88, as for the energy measurement. The shift
observed in the $Z$ resonance peak translates into an uncertainty of 10 MeV in the
determination of the LEP centre-of-mass energy.

\noindent{\bf Background contamination}
\label{backsys}

The expected background level after 
selection cuts is 6\% and the uncertainty on the 
contribution of the different subprocesses
will affect its shape which  may also depend
on the centre-of-mass energy. Since the background is not reweighted,
 this introduces an additional source of  systematic error
both from the background shape 
and its normalization.

\indent{\bf Background shape.}
The normalization uncertainty is estimated independently for each background process.
\begin{itemize}
\item{} The two-photon background, simulated with PHOT02 and PHOJET, is
   normalized to the data in the low visible mass region, M$_{vis}<$ 50 \GeVcsq; the
   difference between the expected and normalized cross-sections 
   is used to estimate the uncertainty of the process, as in~\cite{BEW189VANC}.

\item{} For the $WW$, $ZZ$ and $Zee$ processes the uncertainties are computed from the 
         relative variation of their respective total production cross-sections from the 
        generated energy (183 GeV) down to the fitted energy from the data. This is
        calculated using GENTLE~\cite{GENTLE} for the $WW$ process and PYTHIA for 
        the $ZZ$ and $Zee$ process.  
\end{itemize}

 Each background process is varied according to its uncertainty keeping constant 
the overall background level. The
largest shift observed in each case is quoted in Table~\ref{backshape},
leading to a total systematic error of 8~MeV.

\begin{table}[htb]
\begin{center}
\begin{tabular}{|c|c|}
\hline
Process  & Deviation (MeV) \\
\hline
$WW$ & 4 \\
\hline
$ZZ$ & 2 \\
\hline
$\gamma\gamma$ & 6\\
\hline
$Zee$  & 3  \\
\hline
\hline
Total & 8 \\
\hline
\end{tabular}
\end{center}
\caption[]  
{\protect\footnotesize Systematic error on the centre-of-mass energy 
due to background shape uncertainties. }
\label{backshape}
\end{table}

\indent{\bf Total normalization.}
   The systematic error coming from the  uncertainty in the total background normalization
is estimated by varying all the background processes up and down, according to
their respective uncertainties. The largest  shift 
observed is 31 MeV and is quoted as the systematic error.

\section{Summary and conclusions}
The LEP centre-of-mass energy can be determined 
from the kinematic reconstruction of $q\bar{q}\gamma$ events
with a  $q\bar{q}$ invariant mass around the $Z$ mass.
The average LEP centre-of-mass energy at ALEPH 
for the high energy run of 1997 is measured to be
$$E_{CM} = 182.50 \pm 0.19 ~(stat.) \pm 0.08 ~(syst)~ {\GEV}$$ 
This result is consistent with the estimate from the LEP energy working
group~\cite{lepewg} \\
    $E_{CM}=182.652 \pm 0.050 {\GEV}$.

With the expected increase in statistics at LEP2 and with refined
experimental techniques akin to those used for the $W$ mass
determination~\cite{alephmw183}, the method described here should
provide an alternative measurement of the LEP centre-of-mass energy,
with competitive accuracy. Already, if only this evaluation of the LEP 
centre-of-mass-energy were used in the
$W$ mass determination, the systematic error on $M_W$ coming from the
precision on the beam energy scale would be
90 MeV, ($\Delta M_W/M_W = \Delta E_{CM}/ E_{CM}$),
 which is smaller than the experimental  error of 139 MeV
on $M_W$ from the same data sample~\cite{alephmw183}.

\section{Acknowledgements}
It is a pleasure to congratulate our colleagues 
from the CERN accelerator divisions 
for the successful operation of LEP2. 
We are indebted to the engineers and technicians in all our institutions for 
their contributions to the excellent performance of ALEPH. 
Those of us from non-member 
countries thank CERN for its hospitality and support.
%

%

%
\end{document}